\documentclass[11pt]{iopart}
\usepackage{iopams} 
\usepackage{graphicx}
\usepackage[colorlinks=true,linkcolor=blue,citecolor=red,urlcolor=magenta]{hyperref}
\usepackage[dvipsnames]{xcolor}
\usepackage{charter} 
\usepackage[top=3cm,bottom=3cm,left=2cm,right=2cm,a4paper]{geometry}
  
\begin{document} 

\title[Dynamical properties of a few fermions confined in a double-well potential]{Dynamical properties of a few mass-imbalanced ultra-cold fermions confined in a  
double-well potential}

\author{Dillip K. Nandy \& Tomasz Sowi\'{n}ski}

\address{Institute of Physics, Polish Academy of Sciences \\ Aleja Lotnik\'ow 32/46, PL-02668 Warsaw, Poland}
\ead{nandy@ifpan.edu.pl}

\begin{abstract}
A comprehensive analysis of the exact unitary dynamics of two-component mass-imbalanced fermions in a one-dimensional double-well potential is accomplished by considering the total number of particles maximum up to six. The simultaneous effect of mass imbalance between the flavors and their mutual interactions on the dynamics is scrutinized through the exact diagonalization. In particular, we investigate the occupation dynamics of such systems being initially prepared in experimentally accessible states in which opposite components occupy opposite wells. Moreover, to capture the role of interactions, we also inspect situations in which initial states contain an opposite-spin pair localized in a chosen well. Finally, to assess the amount of quantum correlations produced during the evolution, we analyze the behavior of the von Neumann entanglement entropy between components.
\end{abstract}

\section{Introduction}
 
The monumental progress in the field of laser cooling and trapping of atomic gases has brought unprecedented prospects in the area of quantum engineering. One of them is related to tremendous control on the number of atoms confined in the system \cite{Kohl, 2011SerwaneScience, Blume}. Together with unprecedented tunability of mutual interactions and effective dimensionality, it provides a route to address many theoretical questions on the physical properties of the quantum mesoscopic systems which have so far not been completely understood \cite{2016ZinnerRev,2019SowinskiRPP}. One of them is related to the properties of such few-body systems when they are confined in a double-well potential \cite{2015MurmannPRLb}. Importance of this direction is straightforwardly motivated by an intriguing analogy to the celebrated Josephson effect of coherent tunneling through a classically forbidden region \cite{Josephson,Shockley, Esaki}. Although originally the effect was understood only as an effective and simplified description of the motion of Cooper pairs in superconducting junctions, it rapidly entered the canon of fundamental phenomena triggered by quantum description.

Quantum simulation of the Josephson effect with atomic gases is a very fascinating idea since, besides standard single-particle-like description, one can experimentally validate different models taking into account inter-particle interactions, quantum statistics, internal degrees of freedom, {\it etc.} Initially this scheme was considered in the regime of many particles forming degenerated bosonic gas \cite{Andrews, 1997SmerziPRL, 1997MilburnPRA, 2001MeierPRA, 2004ShinPRL, 2005AlbiezPRL}. In this case, the simplest theoretical description of the effect is served in the framework of the two-mode approximation, {\it i.e.}, by assuming that bosons can occupy only two, appropriately tailored single-particle orbitals describing particles in distinct wells \cite{1997MilburnPRA,2007SalgueiroEPJD}. It is known however that this approach is highly oversimplified whenever inter-particle interactions are relatively stronger or of different than contact character \cite{2009SakmannPRL,2010SakmannPRA,2015GarciaMarchPRA,Dobrzyniecki,roy2020quantum}. 
 
The situation is adequately more challenging when fermionic atoms are considered. From the experimental point of view the most challenging issue, due to an absence of the $s$-wave scattering between indistinguishable fermions, is to cool down the atomic cloud towards the quantum degeneracy regime. The first ground-breaking experiments overcoming this difficulty with so-called sympathetic cooling were performed over 20 years ago \cite{1999DeMarcoScience,2001TruscottScience} and have opened new perspectives for quantum simulators. One of them is the atomtronic experimental realization of the Josephson effect \cite{2012StadlerScience,2015HusmannScience,2015ValtolinaScience,2018BurchiantiPRL}. On the theoretical footing, the description of these kinds of systems is of course much more complicated than bosonic counterparts. It is forced by the Pauli exclusion principle which automatically requires to utilize at least as many single-particle orbitals as the number of particles. 

Recently, the whole story became even more interesting due to the experimental feasibility of simultaneous trapping of fermionic components having different mass elements \cite{Taglieber1, Taglieber2,Wille6Li40K,tiecke2010Feshbach6Li40K,cetina2016ultrafast,Grimm2018DyK}. In this kind of system, the additional parameter that is the mass ratio between the flavors offers an extra degree of freedom which makes the system behave differently than the usual equal mass counterpart \cite{Efimov2, Kartavtsev, Parish, Orso, 2015LoftEPJD,2016PecakNJP,2016DehkharghaniJPhysB,2016PecakPRA}. As argued, it may also significantly influence the topological properties of the system \cite{Wu, Lin, Iskin2}. In addition, one has the experimental flexibility to tune the shape of the trapping potential for particular components. From this perspective, these systems also show some similarity to population imbalanced systems of equal mass flavors \cite{Zwierlein, Partridge}. 

Motivated by these intriguing experiments, here, we theoretically merge the idea of perfect control of systems containing a well-defined number of atoms with the potential importance of degenerated fermionic mixtures of different mass. We aim to investigate the dynamical behavior of the two-component mass-imbalanced few-fermion system confined in a one-dimensional double-well trapping potential. In contrast to recent descriptions in terms of the Multi-Layer Multi-Configurational Time-Dependent Hartree Method \cite{Mistakidis1}, we perform all calculations by a straightforward diagonalization of the many-body Hamiltonian. In this way, we are able to connect different, sometimes counterintuitive, properties of the system with the decomposition of the initial state onto exact many-body eigenstates. Our intention is to shed some new perspective to different problems previously considered for equal-mass fermions \cite{2015BrandtNanoLett,2016SowinskiEPL,Mistakidis2}. First, we want to check the impact of the equal-mass symmetry on the dynamical properties of the system. Second, by considering a specific mixture containing single particle in the selected component we tried to give some dynamical insights towards different small fermi-polaron systems which could be more rigorously studied in the future \cite{2013WenzScience,PhysRevA.96.063603}. Finally, let us also mention that the problem studied may be treated as a starting building-block for further extensions by considering a variety of different external potentials where tunneling plays a crucial role \cite{PhysRevA.83.043604,Chatterjee_2013,Lode2,Richaud}.

For the experimental relevance, we consider the case with a mass ratio corresponding to the mixture of $^{40}$K and $^{6}$Li fermionic atoms. Owing to such practical relevance we show the time dynamics explicitly for this mass ratio in all the system sizes we study here.

\section{The Model}
\begin{figure}
\centering
\includegraphics{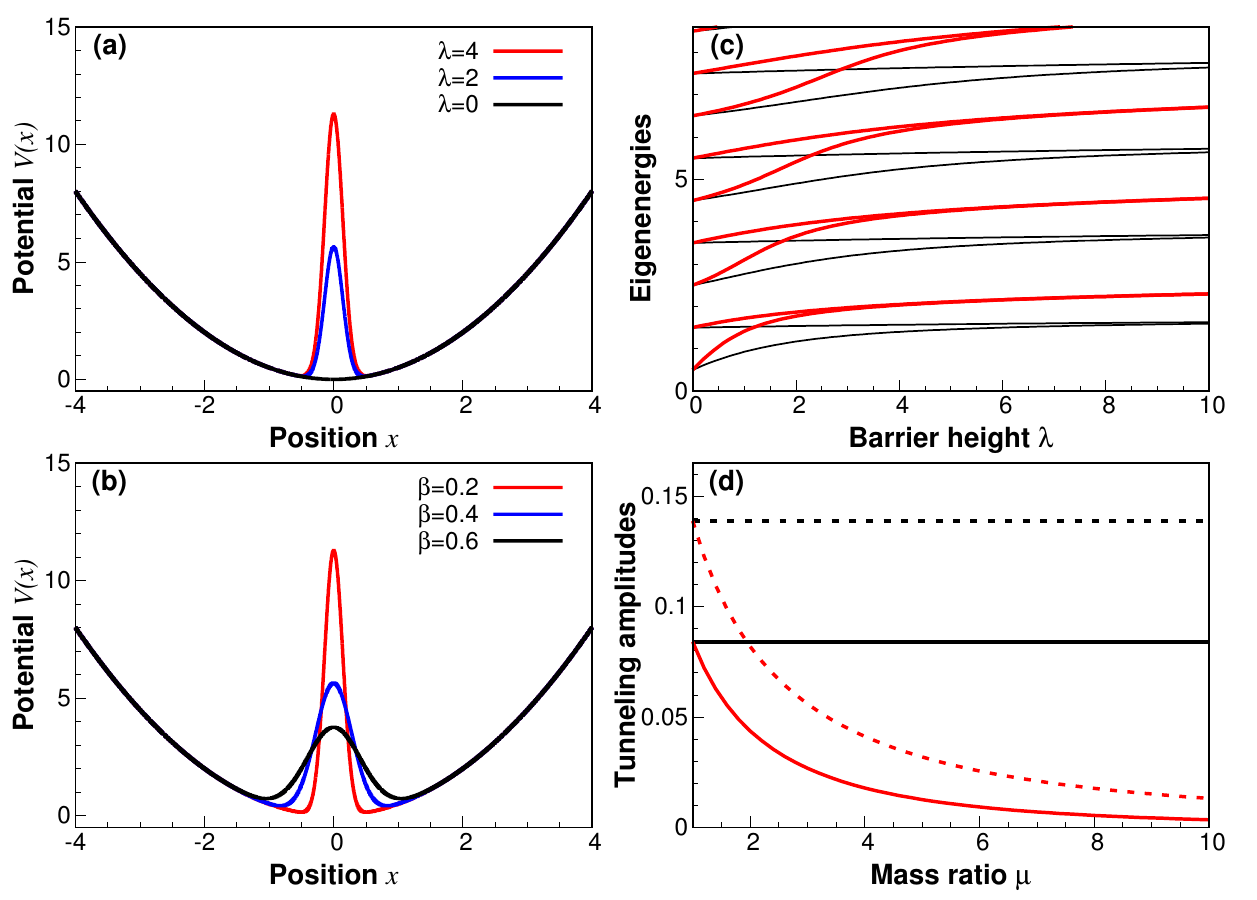}
\caption{Properties of single-particle part of the Hamiltonian. External potential for {\bf (a)} different barrier heights $\lambda$ and fixed $\beta = 0.2$, {\bf (b)} different $\beta$ and fixed barrier height $\lambda = 4.0$. {\bf (c)} Spectra of the single-particle Hamiltonians (\ref{HamSingleA}) and (\ref{HamSingleB}) as functions of $\lambda$ with $\beta = 0.2$ for $\mu = 40/6$. Black thin and red thick lines correspond to spin-$\downarrow$ (light) and spin-$\uparrow$ (heavy) particles, respectively. {\bf (d)} The single-particle tunneling amplitudes of spin-$\uparrow$ (heavy) particles $t_{0,\uparrow}$ (solid red) and $t_{1, \uparrow}$ (dashed red) as functions of mass ratios when $\lambda = 4.0$ and $\beta = 0.2$. For comparison, corresponding tunneling amplitudes for spin-$\downarrow$ (light) particles $t_{0,\downarrow}$ and $t_{1, \downarrow}$ are indicated with horizontal lines. All lengths and energies are expressed in dimensionless units related to a spin-$\downarrow$ (light) particle, $\sqrt{\hbar/m_\downarrow\Omega}$ and $\hbar\Omega$, respectively.} 
\label{Fig1}
\end{figure}

In the present study, we consider a two-component system of a few ultra-cold spinless fermions interacting via a short-range delta-like potential and confined in a one-dimensional external trap having double-well shape. Such a system is described by the Hamiltonian of the form
\begin{eqnarray} \label{Hamiltonian}
 \hat{\mathcal{H}} = \sum_{\sigma} \int\!\!\mathrm{d}x\,\hat{\Psi}^{\dagger}_{\sigma}(x) H_{\sigma} \hat{\Psi}_{\sigma}(x)
+ g \int \mathrm{d}x\,\hat{\Psi}^{\dagger}_{\uparrow}(x) \hat{\Psi}^{\dagger}_{\downarrow}(x) \hat{\Psi}_{\downarrow}(x)  
 \hat{\Psi}_{\uparrow}(x),
\end{eqnarray}
where $\hat{\Psi}_{\sigma}(x)$ is the field operator annihilating fermion of component $\sigma\in\{\downarrow,\uparrow\}$ at position $x$, $g$ is an effective interaction coupling between the flavors, and $H_\sigma$ is a single-particle Hamiltonian describing motion of $\sigma$-component particle. It is clear that the Hamiltonian (\ref{Hamiltonian}) commutes with operators of the number of particles in individual components, $\hat N_\sigma=\int \mathrm{d}x\,\hat{\Psi}^{\dagger}_{\sigma}(x)\hat{\Psi}_{\sigma}(x)$. Therefore in the following, we analyze properties of the system in eigensubspaces of fixed $(\hat{N}_\downarrow,\hat{N}_\uparrow)$. In general, the single-particle Hamiltonians $H_\sigma$ may significantly depend on the component since different species may experience external electromagnetic fields differently. For example, the atomic polarizabilities through the AC Stark shift are typically different and they directly lead to different frequencies of the trapping potential \cite{Cetina}. Having this in mind, here we explore a slightly simplified model assuming that both components experience exactly the same frequency of external trapping. Of course it is still possible to perform more rigorous calculations and consider a more general scenario where trapping frequencies are different. However, the model considered is sufficient to capture the most prominent dynamical properties of these mesoscopic systems. In the mentioned approximation we model the single-particle Hamiltonians as following \cite{Bloch, Mistakidis1, Mistakidis2}
\begin{equation}
H_\sigma = -\frac{\hbar^2}{2m_\sigma}\frac{\mathrm{d}^2}{\mathrm{d}x^2}+\frac{m_\sigma \Omega^2}{2}x^2 
+\frac{\lambda}{\sqrt{\pi}\beta} \mathrm{e}^{-x^2/\beta^2}.
\end{equation}
Thus, both species are confined in a one-dimensional harmonic trap of the same frequency $\Omega$ which is additionally split to two wells by central well of width $\beta$ and strength $\lambda$. In this approximation, parameters $\lambda$ and $\beta$ are independent on the component. This independence is experimentally justified since the parameters are controlled mostly by the intensity and the width of the additional laser beam. If one expresses all quantities in the natural units of the harmonic oscillator related to one of the flavors (in the following we choose the spin-down component) then the single-particle Hamiltonians have a form
\numparts
\begin{eqnarray} 
 H_{\downarrow} &= -\frac{1}{2}\frac{\mathrm{d}^2}{\mathrm{d}x^2} +\frac{1}{2}x^2 
 + \frac{\lambda}{\sqrt{\pi}\beta}\,\mathrm{e}^{-x^2/\beta^2}, \label{HamSingleA} \\
 H_{\uparrow} &= -\frac{1}{2\mu}\frac{\mathrm{d}^2}{\mathrm{d}x^2} +\frac{\mu}{2}x^2 
 + \frac{\lambda}{\sqrt{\pi}\beta}\,\mathrm{e}^{-x^2/\beta^2}, \label{HamSingleB}
\end{eqnarray}
\endnumparts 
where $\mu=m_\uparrow/m_\downarrow$ is the mass ratio between different components atoms. From now, unless stated otherwise, the dynamical properties are discussed explicitly for the $\mu = 40/6$ case corresponding to $^6$Li-$^{40}$K mixture. In Figure~\ref{Fig1}a and Figure~\ref{Fig1}b we present a shape of potential for different parameters while in Figure~\ref{Fig1}c the single-particle spectra for fixed $\beta$ as a function of the barrier height $\lambda$. Black thin and red thick lines correspond to spectra of spin-$\downarrow$ ($m_\downarrow=1$) and spin-$\uparrow$ ($m_\uparrow=40/6$) components, respectively. As expected, for vanishing barrier ($\lambda=0$) harmonic spectrum for both components is reproduced, while in the limit of very deep wells ($\lambda\rightarrow\infty$) two-fold degeneracy between successive even and odd orbitals is obtained.  

For dynamical problems in a double-well potential it is convenient to introduce a basis of states localized in the left ($\psi_{Li}$) and in the right ($\psi_{Ri}$) well of the potential. They can be formed simply by taking the linear combinations of neighboring eigenstates $\phi^{\sigma}_i$ of the single-particle Hamiltonians (\ref{HamSingleA}) and (\ref{HamSingleB}) of the respective flavors:
\numparts
\label{LRorbs}
\begin{eqnarray}
 \psi^{\sigma}_{Li} &=& \frac{1}{\sqrt{2}} (\phi^{\sigma}_{2i} - \phi^{\sigma}_{2i-1}) \label{Lorbs}, \\
 \psi^{\sigma}_{Ri} &=& \frac{1}{\sqrt{2}} (\phi^{\sigma}_{2i} + \phi^{\sigma}_{2i-1}) \label{Rorbs}.
\end{eqnarray}
\endnumparts
One can check that these redefined sets of orbitals also obey the orthonormality condition, however, they are no longer eigenstates of the single-particle Hamiltonian. Using these localized single-particle basis, the fermionic field operators can be expressed as 
\numparts
\begin{eqnarray}
 \hat{\Psi}_{\downarrow}(x) = \sum_i \left [ \psi^{\downarrow}_{Li}(x)\hat{a}_{Li} + \psi^{\downarrow}_{Ri}(x)\hat{a}_{Ri} \right] \\
 \hat{\Psi}_{\uparrow}(x) = \sum_i \left [ \psi^{\uparrow}_{Li}(x)\hat{b}_{Li} + \psi^{\uparrow}_{Ri}(x)\hat{b}_{Ri} \right], 
\end{eqnarray}
\endnumparts
where $\hat{a}$ and $\hat{b}$ are the annihilation operators for the $\downarrow$ and $\uparrow$ fermions, respectively. With the above decomposition of field operators one can now write the Hamiltonian (\ref{Hamiltonian}) as
\begin{eqnarray}
\hat{\mathcal{H}} = \sum_i E_{i\downarrow} \left[\hat{a}^{\dagger}_{Li}\hat{a}_{Li} + \hat{a}^{\dagger}_{Ri}\hat{a}_{Ri} \right]
+ \sum_i E_{i\uparrow} \left[ \hat{b}^{\dagger}_{Li}\hat{b}_{Li} + \hat{b}^{\dagger}_{Ri}\hat{b}_{Ri} \right] \cr
+ \sum_i t_{i\downarrow} \left[\hat{a}^{\dagger}_{Li}\hat{a}_{Ri} + \hat{a}^{\dagger}_{Ri}\hat{a}_{Li} \right] 
+ \sum_i t_{i\uparrow} \left[\hat{b}^{\dagger}_{Li}\hat{b}_{Ri} + \hat{b}^{\dagger}_{Ri}\hat{b}_{Li} \right] \cr
+ g\sum_{ijkl} \sum_{\vec{\alpha}} U^{\vec{\alpha}}_{ijkl} \hat{a}^{\dagger}_{\alpha_{1}i}\hat{b}^{\dagger}_{\alpha_{2}j}
\hat{b}_{\alpha_{3}k}\hat{a}_{\alpha_{4}l}
\label{Ham}
\end{eqnarray}
where $E_{i, \sigma}$ and $t_{i, \sigma}$ can be found as appropriate combinations of single-particle eigenenergies $\varepsilon_{i,\sigma}$ associated with eigenstates $\phi_i^\sigma$
\numparts
\begin{eqnarray}
 E_{i, \sigma} &=& (\varepsilon_{2i+1,\sigma} + \varepsilon_{2i, \sigma})/2, \\
 t_{i, \sigma} &=& (\varepsilon_{2i+1, \sigma} - \varepsilon_{2i, \sigma})/2. 
\end{eqnarray}
\endnumparts
Here, $\vec{\alpha} = (\alpha_1, \alpha_2, \alpha_3, \alpha_4)$ is an algebraic vector of ``left-
right'' indices holding the fact that all four operators come with their own left or right basis state. Interaction energies can be calculated directly from the shape of localized functions 
\begin{eqnarray}
 U^{\vec{\alpha}}_{ijkl} = \int dx \ \psi^*_{\alpha_1i}(x) \ \psi^*_{\alpha_2j}(x) \ \psi_{\alpha_3k}(x) \ \psi_{\alpha_4l}(x).
\end{eqnarray}

Before going to the detailed analysis of the dynamics, we first want to draw attention to the specific behavior of the single-particle tunneling amplitudes as a function of the mass ratios $\mu$. As an example in Figure~\ref{Fig1}d we show this dependence for the single-particle tunneling amplitudes $t_{0,\sigma}$ and $t_{1, \sigma}$. From this plot, it is clear that with the increasing of the mass ratio tunneling amplitudes are monotonically suppressed. Due to this generic behavior of tunneling amplitudes, for each double-well confinement there exists a specific mass ratio for which the tunneling amplitude of a heavier particle in the first excited band $t_{1\uparrow}$ is equal to the tunneling amplitude of a lighter particle in a ground band $t_{0\downarrow}$. This is also the case for higher bands but obviously happens for different mass ratios. In the following, we aim to examine also the consequences of this specific resonance condition and its influence on the dynamical properties of the system. 

\section{The Method}
The dynamical behavior of the system is studied by solving the time-dependent many-body Schr\"{o}dinger equation. In our numerical approach, first, we fix the number of particles in the respective components $(N_\downarrow,N_\uparrow)$ and then we construct the matrix form of the many-body Hamiltonian (\ref{Ham}) in the Fock basis by using the redefined single-particle orbitals (\ref{LRorbs}). In the next step, for a given interaction $g$, we do the exact diagonalization of this constructed Hamiltonian to obtain many-body eigenstates $|E_n \rangle$ with corresponding eigenenergies $E_n$. As the Hilbert space size grows exponentially with the number of orbitals, we truncate the single-particle orbitals on a reasonable cut-off value which of course depends on the system's size and also on the interaction strength. Nevertheless, for the interaction range, we are interested here, we find that for most of the system size the results are hardly affected by the inclusion of higher orbitals than the cutoff values.

Here, we assume that the evolution is unitary and it is governed solely by the many-body Hamiltonian (\ref{Hamiltonian}). Therefore, having the system initially prepared in the state $|\mathtt{ini} \rangle$ the state of the system at later moment $t$ is given by
\begin{equation} \label{Cndefin}
|\Phi(t) \rangle = \sum_n e^{-iE_nt} C_n |E_n \rangle,
 \label{dynPsi}
\end{equation}
where the coefficients $C_n = \langle E_n| \mathtt{ini} \rangle$. In principle, the summation index can run over full Hilbert space dimension, however it is always crucial, especially for large Hilbert space dimension, to find a suitable number of eigenstates that can yield the desired results. In practice, this can be achieved by determining the initial state fidelity $F=|\langle\Phi(0)|\mathtt{ini} \rangle|^2$. Here, we ensure that the fidelity is always larger than $99\%$ in all the cases. Having the state of the system at any moment $|\Phi(t)\rangle$ we can determine different observables which are accessible experimentally. The simplest are related to the single-particle density profiles of the form
\begin{eqnarray}
 \rho_{\sigma}(x,t) = \langle \Phi(t)| \hat{\Psi}^{\dagger}_{\sigma}(x) \hat{\Psi}_{\sigma}(x) |\Phi(t) \rangle.
 \label{dynRho}
\end{eqnarray}
Although the profiles are well understood quantities, the macroscopic flow between the wells is described much better in the language of integrated quantities. Therefore, for further convenience we introduce the occupation numbers in the left and right well as
\begin{equation} \label{occupations}
 N_{L\sigma}(t) = \int^0_{-\infty}\!\mathrm{d}x\, \rho_{\sigma}(x,t), \,\, 
 N_{R\sigma}(t) = \int^{\infty}_0\!\mathrm{d}x\, \rho_{\sigma}(x,t).
\end{equation}

\begin{figure}
\centering
\includegraphics[scale=0.42]{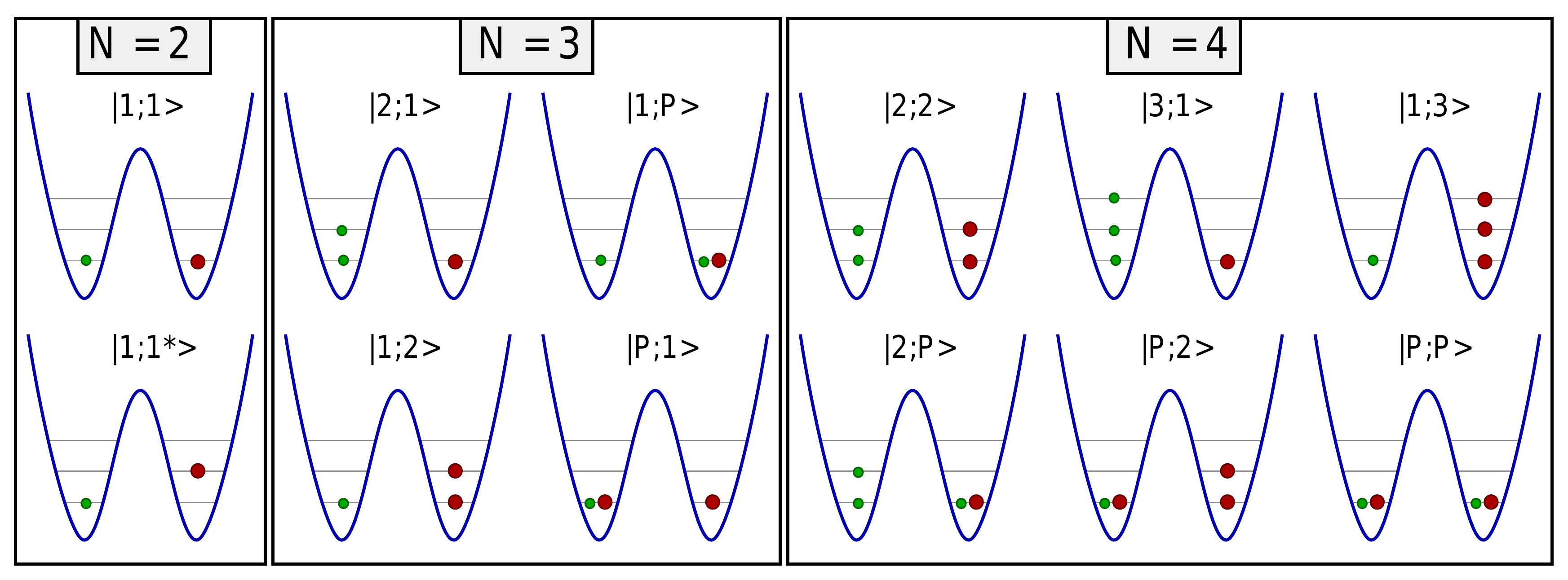}
\caption{Pictorial representation of different initial configurations of $N=N_\downarrow+N_\uparrow=\{2,3,4\}$ particles considered in the main text. The small green and big red dots correspond to spin-$\downarrow$ (lighter) and spin-$\uparrow$ (heavier) particles, respectively. Horizontal lines drawn in each well refer to single-particle orbitals of the single-particle Hamiltonian.} 
\label{Fig2}
\end{figure}

\section{Two-particle system} \label{Sec4}

We start a whole discussion with the simplest case of two opposite-spin fermions. In this case, we consider two different initial states. First, we assume that particles initially occupy opposite wells in the ground band, {\it i.e.}, the initial state has a form $|{\bf 1};{\bf 1}\rangle = a^{\dagger}_{L0} b^{\dagger}_{R0} |\mathtt{vac}\rangle$. Notice that our notation exploits directly the left-right mirror symmetry of the problem and assume that left (right) well is occupied by spin-$\downarrow$ (spin-$\uparrow$) particles. In the second case we relax the condition of minimum energy configuration and we consider the scenario where the heavier particle is singly excited in its well, $|{\bf 1};{\bf 1}^*\rangle=a^{\dagger}_{L0} b^{\dagger}_{R1} |\mathtt{vac}\rangle$ (symbol ``$*$'' means a single spatial excitation in a well). In this way, we are able to examine the behavior of the system when we cross the resonance condition for tunneling amplitudes mentioned above. Pictorially, both initial states are presented in the left panel in Figure~\ref{Fig2}. 

\begin{figure}
\centering
\includegraphics{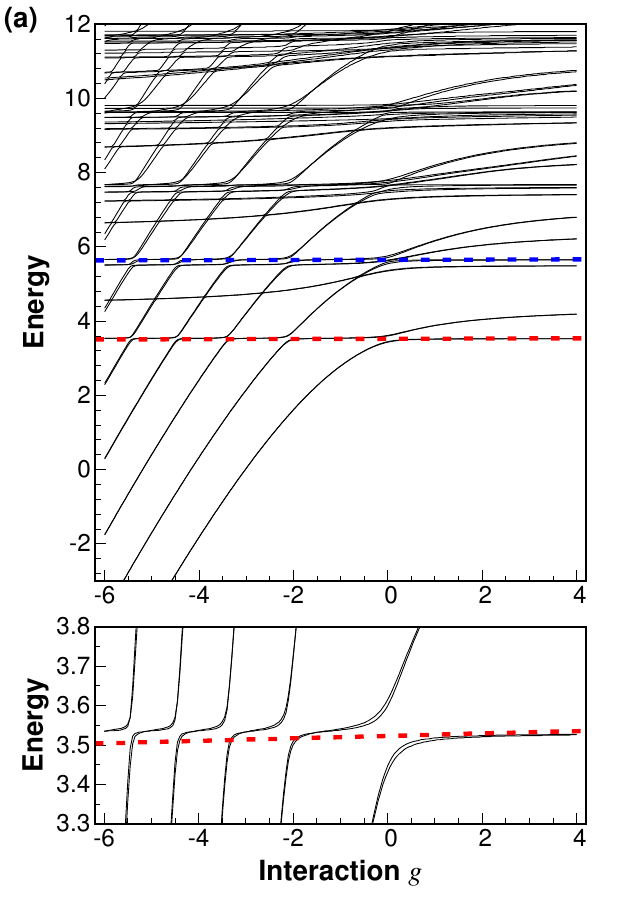}
\includegraphics{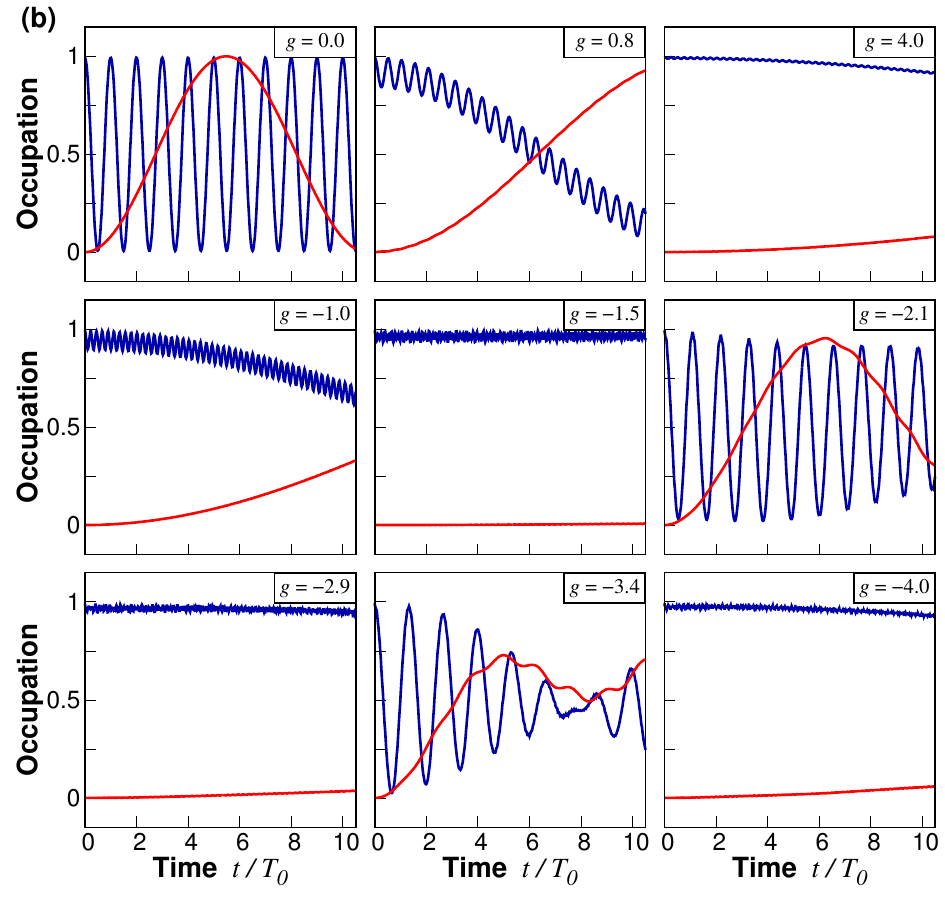}
\includegraphics{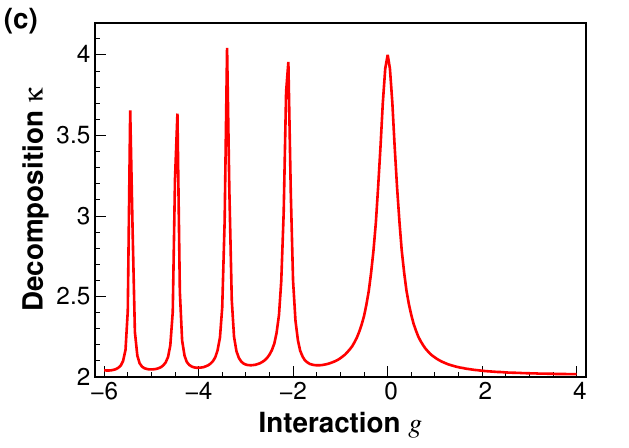}
\includegraphics{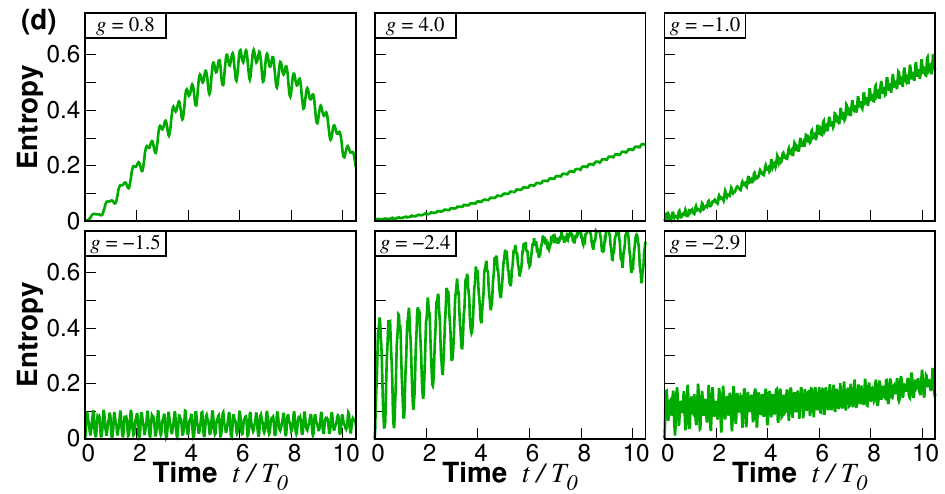}
\caption{Dynamical properties of the system with $N_\downarrow=N_\uparrow=1$ particles for $\mu=40/6$ confined in the double-well trap with $\beta=0.2$ and $\lambda=4.0$. {\bf (a)} The many-body spectrum of the Hamiltonian (\ref{Hamiltonian}). The dashed red and blue lines correspond to energies of the initial states $|{\bf 1};{\bf 1}\rangle$ and $|{\bf 1};{\bf 1}^*\rangle$, respectively. The bottom plot shows magnification of the spectrum close to the energy of the initial state $|{\bf 1};{\bf 1}\rangle$ and explains emergence of resonant behavior in the dynamics. {\bf (b)} Left-well occupations $N_{L\sigma}(t)$ as functions of time for different interactions for a system initially prepared in the state $|{\bf 1};{\bf 1}\rangle$. The blue and red line refer to the lighter and heavier component, respectively. Note, enhancement of tunneling close to resonance points captured by the decomposition number $\kappa$ in the plot (c). {\bf (c)} Effective number of many-body eigenstates $\kappa$ contributing to the initial state $|{\bf 1};{\bf 1}\rangle$ as a function of interactions. Note, significant increase of the parameter at resonant interaction strengths. {\bf (d)} Evolution of the inter-component entanglement entropy $S(t)$ for different interactions. All energies and interaction strengths $g$ are expressed in dimensionless units related to a spin-$\downarrow$ (light) particle, $\hbar\Omega$ and $(\hbar^3\Omega/m_\downarrow)^{1/2}$, respectively. Time is normalized to the oscillation period of a free spin-$\downarrow$ (light) particle in the lowest band, $T_0=\pi\hbar/t_{0\downarrow}$. \label{Fig3}}
\end{figure}

It is instructive to start the analysis by plotting the many-body spectrum as a function of interaction $g$ and compare it with the average energy in the initial states. Since in both initial states considered particles are spatially separated, the average energy $\langle \mathtt{ini}|\hat{\cal H}|\mathtt{ini}\rangle$ almost does not depend on interaction strength $g$. In contrast, the many-body spectrum is interaction sensitive. By direct comparison of these energies (see Figure~\ref{Fig3}a) one can notice that for some particular interaction strengths on the attractive branch, possible energy matching occurs. Since evolution is an energy-conserving process, we may suspect that for these specific interactions some changes of the dynamical properties of the system may be present.

To get a better view of the dynamical properties of the system we focus on the time evolution of occupations (\ref{occupations}) (see Figure~\ref{Fig3}b). Since initially both particles occupy the lowest band, in the case of non-interacting system the dynamics is governed solely by $t_{0\downarrow}$ and $t_{0\uparrow}$ tunneling amplitudes. When interactions increase towards repulsions or attractions we observe significant suppression of the tunneling. This fact is a direct manifestation of energy conservation -- single-particle tunnelings lead to states having both particles occupying the same well. Many-body energies of these states however, due to the interaction energy, do not match the initial energy of the system. Only in the second-order process the state with the same energy having particles interchanged may be reached. This process however has the effective amplitude suppressed by interactions, $\sim t_{0\downarrow}t_{0\uparrow}/|g|$. For repulsive interactions, this phenomenological argumentation is always valid since interactions cannot connect the initial state to any other many-body state having substantially different many-body energy. The situation is a little bit different for attractive forces. As clearly visible on the many-body spectrum, for some particular interaction strengths, highly excited many-body states may have exactly the same energy as the energy of the initial state. Since couplings to these states are provided by interaction terms, energy-conservative evolution may be much richer. Our numerical calculations fully support this observation. As is seen in Figure~\ref{Fig3}b for some particularly chosen interactions ($g\approx -2.4$, $g\approx -3.6$, {\it etc.}) the tunneling is enhanced and particles freely exchange their wells. We checked that a very similar mechanism is present for the second initial state $|{\bf 1};{\bf 1}^*\rangle$. Although single-particle tunneling for the excited particle is different, interactions play exactly the same role -- they suppress density flow when couplings to other many-body states violate conservation of the energy and they enhance collective tunneling whenever other many-body states are accessible without this violation. Importantly and surprisingly, our numerical calculations show that no significant change of this behavior is present when the resonant condition between tunnelings ($t_{0\downarrow}=t_{1\uparrow}$) is reached for the initial state $|{\bf 1};{\bf 1}^*\rangle$. 

To make the analysis more comprehensive, we also examined the artificial model with single-particle tunneling terms completely switched off. It turns out that, although the single-particle terms play a dominant role far from these resonant points, their importance is negligible at resonances where a whole dynamics is governed almost solely by interaction terms.

The suppression or enhancement of the density flow through the barrier can be also explained by inspecting the decomposition of the initial state to the many-body eigenstates of the Hamiltonian. In fact, the dynamics of the system is governed only by a few eigenstates dominating in the initial state. To quantify this number, we analyze the properties of the decomposition coefficients $C_n $ defined in (\ref{Cndefin}) by calculating the inverse participation ratio 
\begin{equation}
 \kappa = \frac{1}{\sum_{n} |C_{n}|^4}.
\end{equation}
The quantity $\kappa$ essentially determines an effective number of many-body eigenstates taking part in the decomposition of an initial state. As shown in the Figure~\ref{Fig3}c, whenever interaction strength $g$ is close to the resonant value the number $\kappa$ significantly increases. It simply means, a larger number of states take part in determining the dynamics of the system and substantially change its properties. In particular, it is interesting to find this decomposition for interactions for which the density flow is significantly restrained. As we obtained from the decomposition, the effective number of eigenstates contributing to the dynamics is around $\kappa\approx 2$, {\it i.e.}, only two eigenstates provide a whole dynamics of the system. As a matter of fact, the Fock-space decomposition of these two eigenstates (say $|E_1 \rangle$ and $|E_2 \rangle$) is very simple -- they can be almost perfectly expressed as a linear combination of the initial state and its respective mirror reflection (the state with interchanged wells). Hence, as a consequence of very strong interactions, a description of the whole system is significantly simplified and the dynamics of the many-body system can be viewed as a simple two-level system. We checked that this mechanism is generic for any initial state provided that interactions strongly suppress the density flow.

At this point it should be mentioned that even in the cases when density flow between wells is strongly suppressed by interactions the many-body state of the system continuously evolves. It can be clearly visible when correlations between particles are considered. One of the most natural measures of quantum correlations present in the system is the inter-component entanglement entropy defined as
\begin{equation}
S(t) = -\mathrm{Tr}_\uparrow \left[\hat\rho_\downarrow(t) \ln\hat \rho_\downarrow(t)\right]=-\mathrm{Tr}_\downarrow \left[\hat\rho_\uparrow(t) \ln\hat \rho_\uparrow(t)\right],
\end{equation}
where the reduced density matrix of a chosen component $\sigma$ can be calculated by tracing out opposite component from the projector to a temporal state of the system, $\hat\rho_\sigma(t)=\mathrm{Tr}_{\sigma'} \left[|\Phi(t)\rangle\langle\Phi(t)|\right]$.

In Figure~\ref{Fig3}d we display time evolution of the entropy for different interactions. Initially, particles occupying opposite wells are uncorrelated and the entropy vanishes. It is clear that even in the cases when density flow is frozen and particles remain in their initial wells, the entropy fluctuates signaling evolution of the state of the system. Of course in situations when the flow of particles is evident, entanglement entropy is developed more significantly. 

\begin{figure} 
\centering 
\includegraphics{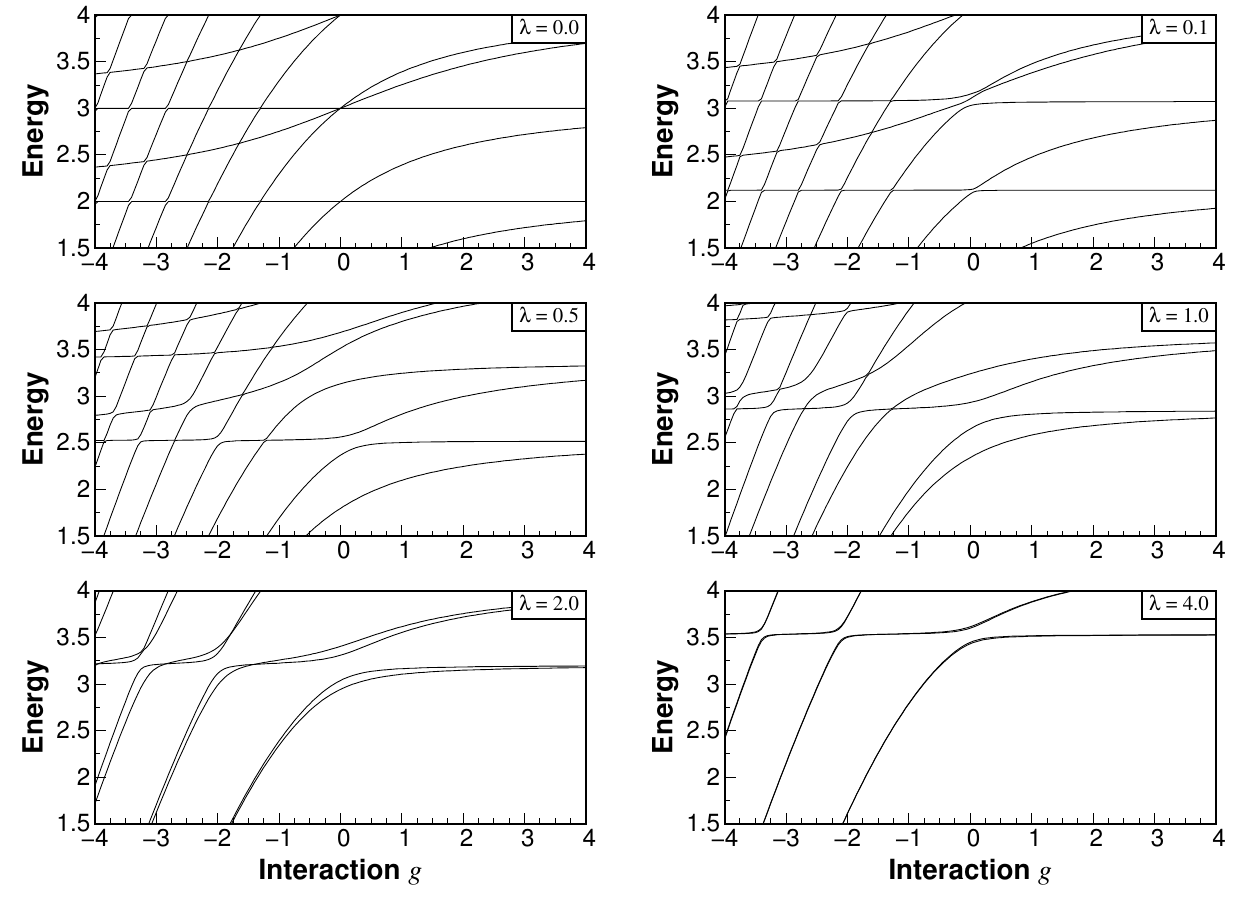}
\caption{Evolution of the many-body spectrum of the system of $N_\downarrow=N_\uparrow=1$ particles when the barrier height $\lambda$ is increased. One observes that along with increasing $\lambda$, first avoided crossings appear in the spectrum. Then, for sufficiently deep wells, double quasi-degeneracy of many-body states is formed. In all plots energies and interaction strengths are measured in natural units of the problem $\hbar\Omega$ and $(\hbar^3\Omega/m_\downarrow)^{1/2}$, respectively. \label{Fig4}}
\end{figure}
Finally, let us visualize how the avoided crossings of quasi-degenerated energy levels appear in the many-body spectrum of the system. In Fig.~\ref{Fig4} we show the evolution of the many-body spectrum of the system with $N_\downarrow=N_\uparrow=1$ particles when the barrier height $\lambda$ is increased. First, for $\lambda = 0.0$ one recognizes the well-known spectrum of the system confined in a harmonic trap with no avoided crossings. Then, along with increasing anharmonicity ($\lambda = 0.1, 0.5, 1.0$) the avoided crossings appear and they are successively enhanced. At these values of anharmonicity parameter, the appearance of the avoided crossings is a direct manifestation of emerging couplings between the center-of-mass and internal (relative motion) degrees of freedom triggered by the anharmonicity of the confining potential. For larger anharmonicity ({\it e.g.} $\lambda = 2.0, 4.0$) another effect of the double-well potential (quasi-degeneracy of single-particle orbitals) starts to play a significant role and one observes how the double quasi-degeneracies between energy levels are formed. For the quite deep wells considered in our work ($\lambda = 4$) the resonance behavior is therefore attributed to the simultaneous effect of avoided crossings and the single-particle degeneracies of the energy levels forced respectively by anharmonicity and single-particle degeneracy of the double-well confinement.

\section{Three-particle system}
Now let us investigate systems with $N=N_\downarrow+N_\uparrow=3$ particles. These are the minimal systems for which the quantum statistics between indistinguishable particles has to be taken into account. In the following, first we focus in cases when both flavors are initially prepared in the distinct wells. Depending on the distribution of particles between flavours, we distinguish two initial states of this kind, $|{\bf 2};{\bf 1}\rangle = a^{\dagger}_{L0} a^{\dagger}_{L1} b^{\dagger}_{R0} |\mathtt{vac}\rangle$ and $|{\bf 1};{\bf 2}\rangle = a^{\dagger}_{L0} b^{\dagger}_{R0} b^{\dagger}_{R1} |\mathtt{vac}\rangle$. Moreover, to make the discussion more general, we additionally consider two other initial states in which opposite-spin particles are paired in a chosen well. These states can be written as $|{\bf 1};{\bf P}\rangle = a^{\dagger}_{L0} a^{\dagger}_{R0} b^{\dagger}_{R0} |\mathtt{vac}\rangle$ and $|{\bf P};{\bf 1}\rangle = a^{\dagger}_{L0} b^{\dagger}_{L0} b^{\dagger}_{R0} |\mathtt{vac}\rangle$, where the symbol ``${\bf P}$'' stands for a pair of opposite-spin particles occupying the lowest orbital in a well. Note that these states are essentially different since the remaining (unpaired) particle belongs to lighter or heavier component. The notation used is consistent with the one used previously, {\it i.e.}, numbers associated to the left and right well correspond to $\downarrow$- and $\uparrow$-particles, respectively. Schematic representations of all these three-particle states are displayed in the middle panel in Figure~\ref{Fig2}.

\begin{figure}
\centering
\includegraphics{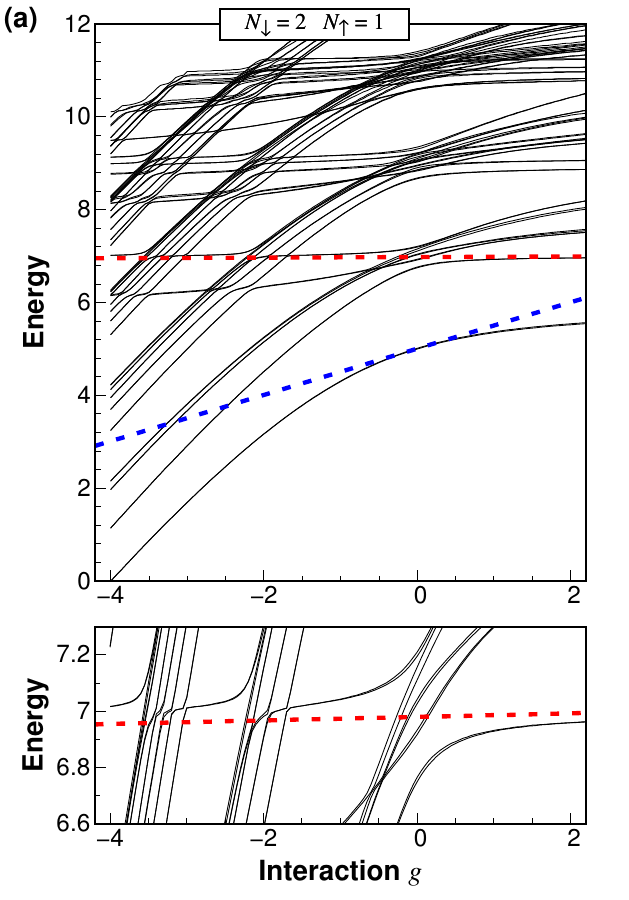}
\includegraphics{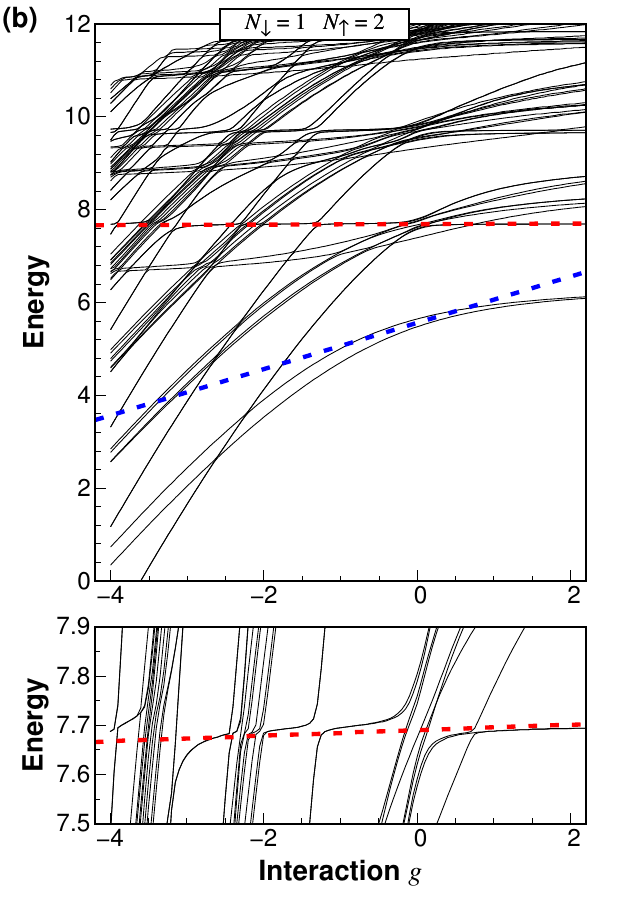}
\includegraphics{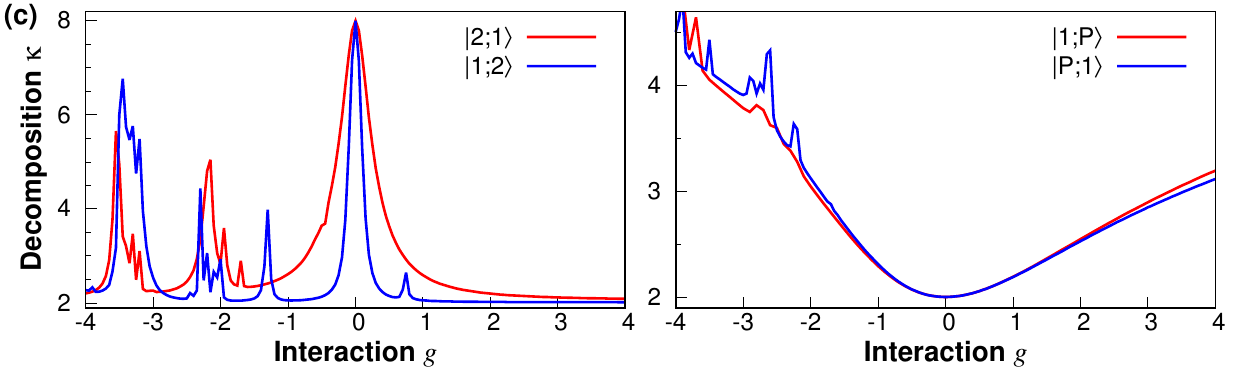}
\includegraphics{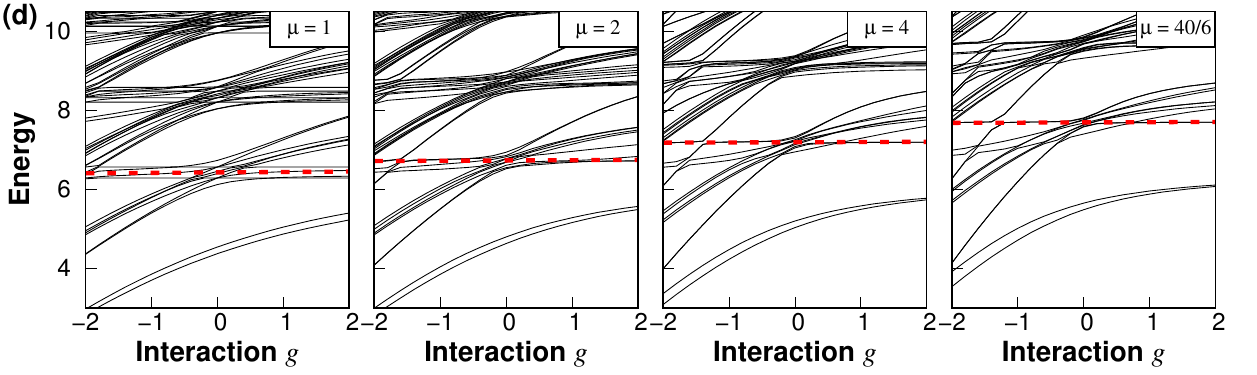}
\caption{Dynamical properties of the system with $N_\downarrow+N_\uparrow=3$ particles for $\mu=40/6$ confined in the double-well trap with $\beta=0.2$ and $\lambda=4.0$. {\bf (a-b)} The many-body spectra of the Hamiltonian (\ref{Hamiltonian}) for different distribution of particles. The thick dashed red (blue) lines correspond to the energies of initial states $|{\bf 2};{\bf 1}\rangle$ and $|{\bf 1};{\bf 2}\rangle$ ($|{\bf 1};{\bf P}\rangle$ and $|{\bf P};{\bf 1}\rangle$), respectively. The bottom plot shows magnification of the spectra close to the energy of the initial states $|{\bf 1};{\bf 2}\rangle$ and $|{\bf 2};{\bf 1}\rangle$, respectively. {\bf (c)} Effective number of many-body eigenstates $\kappa$ contributing to different initial states as a function of interactions. Note a significant difference between states without (left panel) and with (right panel) paired particles. {\bf (d)} Many-body spectrum of the system with $(N_{\downarrow},N_{\uparrow})=(1,2)$ particles obtained for different mass ratio $\mu$ plotted in the vicinity of the initial state $|{\bf 1};{\bf 2}\rangle$. It shows the effect of mass imbalance on the many-body spectra and explains emergence of resonant behavior of the tunneling. See the main text for explanation. All energies and interaction strengths $g$ are expressed in dimensionless units related to a spin-$\downarrow$ (light) particle, $\hbar\Omega$ and $(\hbar^3\Omega/m_\downarrow)^{1/2}$, respectively. \label{Fig5}}
\end{figure}

The energy spectra for both distributions of particles $(N_{\downarrow},N_\uparrow) = (2,1)$ and $(N_{\downarrow},N_\uparrow) = (1,2)$ are shown in Figure~\ref{Fig5}a. On top of them the average energies of corresponding initial states are plotted. Exactly as in the case of two-particle system we find that energies of the initial states $|{\bf 2};{\bf 1}\rangle$ and $|{\bf 1};{\bf 2}\rangle$ (shown as thick dashed red lines) are almost insensitive to the interaction strength. Contrary, the two other states $|{\bf 1};{\bf P}\rangle$ and $|{\bf P};{\bf 1}\rangle$ containing paired particles (thick dashed blue lines) have energies obviously dependent on interactions.

Although the states $|{\bf 2};{\bf 1}\rangle$ and $|{\bf 1};{\bf 2}\rangle$ are essentially different, it turns out that their dynamical properties are qualitatively the same. This similarity can be explained by comparing relations between their initial energies and underlying many-body spectra of corresponding systems (see magnifications of the spectra in Fig.~\ref{Fig5}a and Fig.~\ref{Fig5}b). On a quantitative level, this comparison shows that both the states have very similar interaction dependence of the effective number of contributing states $\kappa$ (left panel in Fig.~\ref{Fig5}c). Consequently, exactly as in the two-particle case, one distinguishes two different scenarios. {\bf (I)} The energy of the initial state evidently {\it mismatch} with the energy of any of the eigenstates of the system. Then, the decomposition of the initial state $\kappa\approx 2$ and consequently the tunneling dynamics is suppressed. {\bf (II)} The initial energy is close to some eigenenergies of the system and the initial state can be decomposed to a larger number of eigenstates ($\kappa\gg 2$). In this case, the tunneling through the barrier is constructively supported by interactions and evident speed-up in tunneling is visible. At this point however it should be emphasized that, in contrast to cases with $N=2$ particles, for stronger attractions, the initial state energy becomes immersed in the plethora of many-body eigenstates having close energies which contribute to the dynamics (note rapid and unstable oscillations of the decomposition number $\kappa$). In consequence, in this range of interactions, the dynamics become very sensitive to the tuning of the interaction strength and in practice it is unpredictable.

Situation is significantly different when the initial states $|{\bf 1};{\bf P}\rangle$ and $|{\bf P};{\bf 1}\rangle$ are considered. As indicated previously, the energy of these initial states depends on the interaction strength $g$. Moreover, the quantum statistics starts to play an important role in establishing the dynamical properties of the system. It is quite clear in the limit of vanishing interactions -- the Pauli exclusion principle prohibits single-particle tunneling for one of the flavors and the dynamics is present only for the component having single particle. In this limit, the initial state is decomposed exactly to two eigenstates of the system ($\kappa=2$). For increasing interactions (repulsive as well as attractive) we notice a monotonic increase of the number of states contributing to the initial state (right panel in Fig.~\ref{Fig5}c). Thus the system does not display any resonant behavior with respect to varying interactions. 

The emergence of regions of a strong suppression in tunneling can also be well understood when the response of the many-body energy spectra to the mass ratio $\mu$ is analyzed. As an example, in Figure~\ref{Fig5} we display situation for system with $(N_\downarrow,N_\uparrow)=(1,2)$. It is clearly noticeable, specifically at the attractive part that for the equal mass system the energy of the initial state $|{\bf 1};{\bf 2}\rangle$ is continuously embedded in some of eigenenergies of the system. As a result, the system favors the resonant situation for any interaction. Thereafter, with the increase of the mass ratio, the eigenenergy levels are more repelled from the initial-state energy and the initial state coincides only with one of the eigenstates of the many-body Hamiltonian. Such localization of the initial state simply implies the reason for the suppressed dynamics. Only for specific interaction strengths, other eigenstates of the system become close to the initial state and then the resonant tunneling is supported. Analogously, one can apply similar argumentation to the system with $(N_{\downarrow},N_\uparrow)=(2,1)$ particles with akin conclusions.  

\section{Four-particle system}
In this section, we discuss dynamical properties of four-particle systems. Of course, along with increasing number of particles, possible number of interesting initial configurations also increases. To uncover generic roles played by interactions and the quantum statistics we focus on six the most interesting representatives. First three states correspond to the initially separated flavors and depending on the particle imbalance between flavors they form:
$|{\bf 2};{\bf 2}\rangle=\hat{a}_{L0}^\dagger\hat{a}_{L1}^\dagger\hat{b}_{R0}^\dagger\hat{b}_{R1}^\dagger|\mathtt{vac}\rangle$, 
$|{\bf 3};{\bf 1}\rangle=\hat{a}_{L0}^\dagger\hat{a}_{L1}^\dagger\hat{a}_{L2}^\dagger\hat{b}_{R0}^\dagger|\mathtt{vac}\rangle$, and
$|{\bf 1};{\bf 3}\rangle=\hat{a}_{L0}^\dagger\hat{b}_{R0}^\dagger\hat{a}_{R1}^\dagger\hat{b}_{R2}^\dagger|\mathtt{vac}\rangle$. Next two states have opposite-spin particles paired in a chosen well: $|{\bf 2};{\bf P}\rangle=\hat{a}_{L0}^\dagger\hat{a}_{L1}^\dagger\hat{a}_{R0}^\dagger\hat{b}_{R0}^\dagger|\mathtt{vac}\rangle$ and $|{\bf P};{\bf 2}\rangle=\hat{a}_{L0}^\dagger\hat{b}_{L0}^\dagger\hat{b}_{R0}^\dagger\hat{b}_{R1}^\dagger|\mathtt{vac}\rangle$. Finally, the last initial state we consider has two pairs of fermions prepared in opposite wells, $|{\bf P};{\bf P}\rangle=\hat{a}_{L0}^\dagger\hat{a}_{R0}^\dagger\hat{b}_{L0}^\dagger\hat{b}_{R0}^\dagger|\mathtt{vac}\rangle$. All these states are pictorially presented in the right panel in Figure~\ref{Fig2}.

\begin{figure}
\centering
\includegraphics{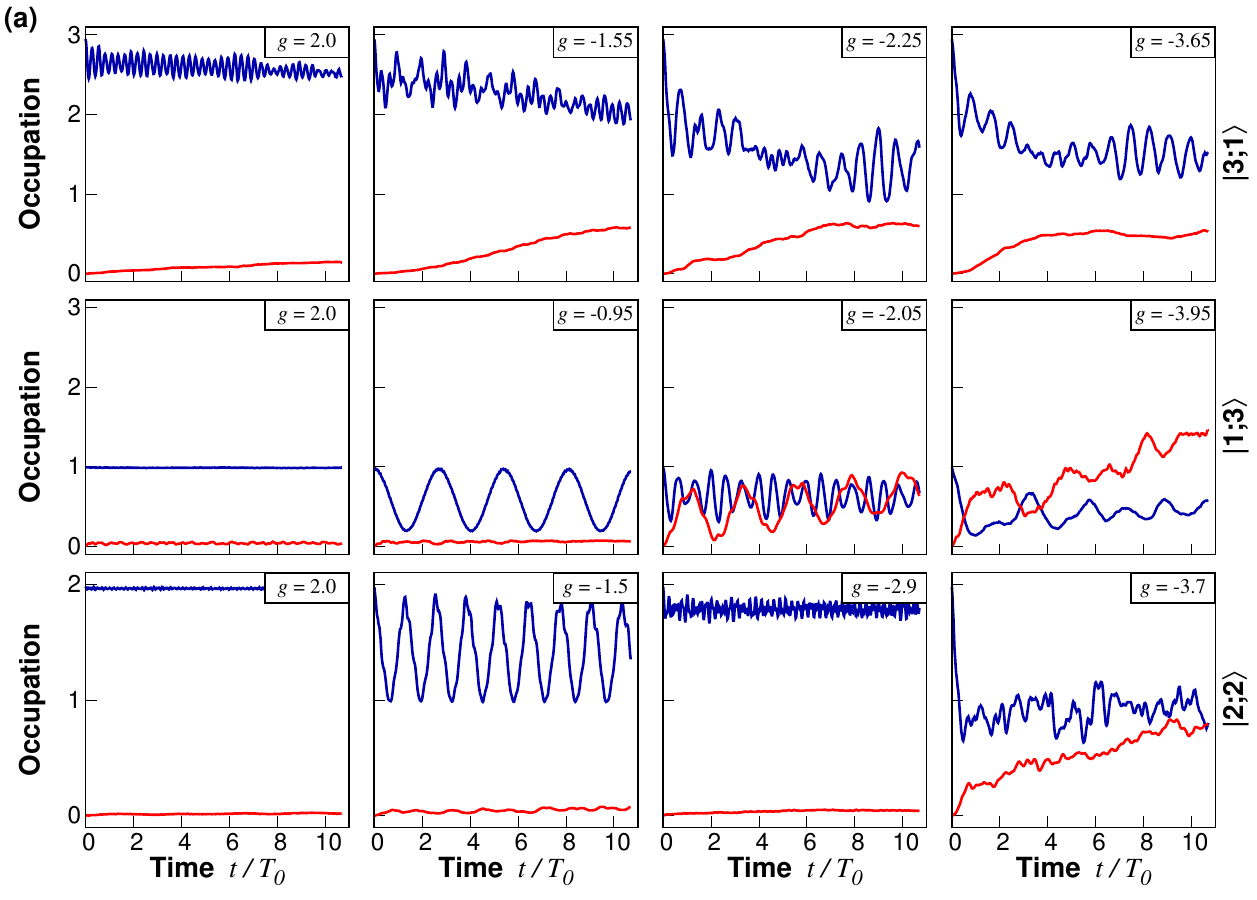}
\includegraphics{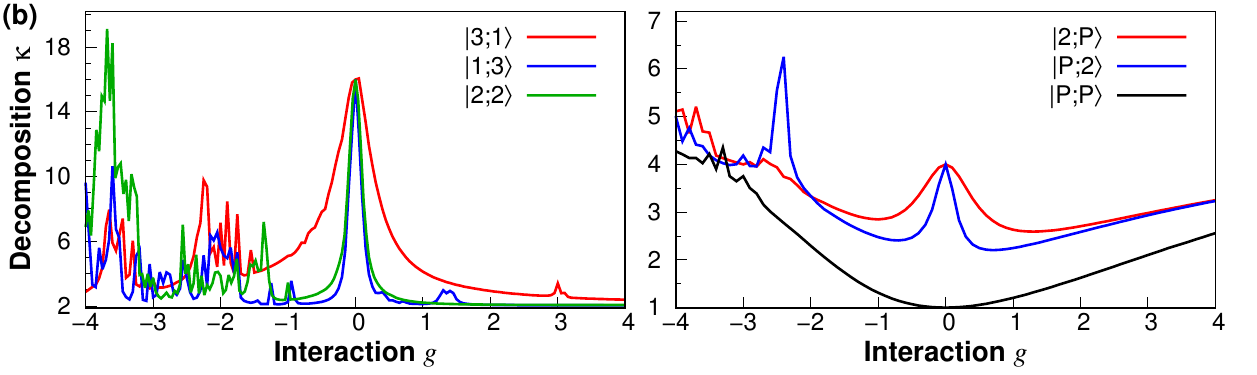} 
\caption{Dynamical properties of the system with $N_\downarrow+N_\uparrow=4$ particles and $\mu=40/6$ confined in the double-well trap with $\beta=0.2$ and $\lambda=4.0$. {\bf (a)} Left-well occupations $N_{L\sigma}(t)$ as functions of time for different interactions for the system initially prepared in the states of ideally separated components $|{\bf 3};{\bf 1}\rangle$, $|{\bf 1};{\bf 3}\rangle$, and $|{\bf 2};{\bf 2}\rangle$. The blue and red line refer to the lighter and heavier component, respectively. {\bf (b)} Effective number of many-body eigenstates $\kappa$ contributing to different initial states as a function of interactions. Interaction strength $g$ is expressed in dimensionless units related to a spin-$\downarrow$ (light) particle, $(\hbar^3\Omega/m_\downarrow)^{1/2}$. Time is normalized to the oscillation period of free spin-$\downarrow$ (light) particle in the lowest band, $T_0=\pi\hbar/t_{0\downarrow}$. \label{Fig6}}
\end{figure}

\begin{figure}
\centering
\includegraphics[scale=0.15]{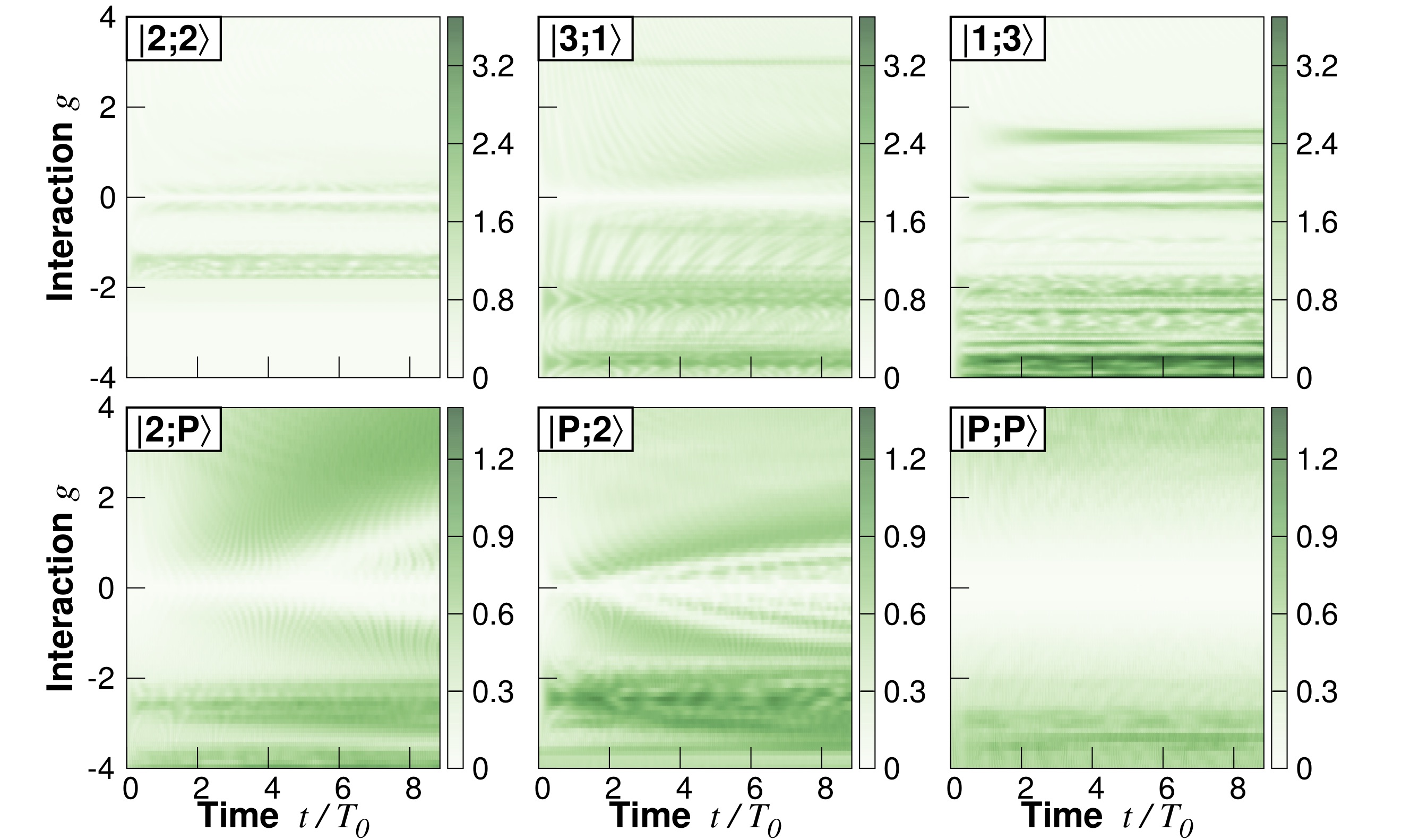}
\caption{Evolution of the inter-component entanglement entropy $S$ as a function of interactions for different initial states containing $N_\downarrow+N_\uparrow=4$ particles. Note significant difference between evolution of initially separated components ($|{\bf 2};{\bf 2}\rangle$, $|{\bf 3};{\bf 1}\rangle$, $|{\bf 1};{\bf 3}\rangle$) and evolution of the system when initially opposite-spin fermions pair is localized in selected well ($|{\bf 2};{\bf P}\rangle$, $|{\bf P};{\bf 2}\rangle$, $|{\bf P};{\bf P}\rangle$). Interaction strength $g$ is expressed in dimensionless units related to a spin-$\downarrow$ (light) particle, $(\hbar^3\Omega/m_\downarrow)^{1/2}$. Time is normalized to the oscillation period of free spin-$\downarrow$ (light) particle in the lowest band, $T_0=\pi\hbar/t_{0\downarrow}$.\label{Fig7}}
\end{figure}

Similarly to the case of $N=3$ particles discussed before, the dynamical properties of systems prepared initially in states $|{\bf 3};{\bf 1}\rangle$ and $|{\bf 1};{\bf 3}\rangle$ are qualitatively comparable. In Figure~\ref{Fig6}a we show the time evolution of left-well occupations for some particularly chosen interactions. It is clearly visible that the general scheme is still valid, {\it i.e.} interactions, due to the conservation of the energy, typically suppress tunnelings between wells. However, on the attractive branch, for some particular values of interactions, the tunneling is strongly enhanced. We checked that this kind of behavior is also present in the balanced case represented by the initial state $|{\bf 2};{\bf 2}\rangle$ which has no counterpart in the $N=3$ systems. The physical picture outlined is in full agreement with the dependence of the decomposition number $\kappa$ displayed in the left panel in Figure~\ref{Fig6}b. It is clear, that for all three states of initially separated clouds the decomposition coefficients behave similarly -- they decrease with increasing interactions except the resonant points where a larger number of many-body states contribute to the chosen initial state. For stronger attractions, regions of unpredictable dynamics (indicated by rapid oscillations of $\kappa$) also occur. Comparison to systems with $N=3$ particles suggests that this kind of dynamical response of the system could be the general property for any number of particles, as long as opposite-spin components are initially separated and well-localized in opposite wells. It should be also pointed that for relatively weak interactions, {\it i.e.}, far from the region of unpredictable dynamics, decomposition coefficient $\kappa$ for the initial state $|{\bf 3};{\bf 1}\rangle$ decreases much slower than for states $|{\bf 1};{\bf 3}\rangle$ and $|{\bf 2};{\bf 2}\rangle$. It means, in this case, to suppress the tunneling one needs stronger interactions. This observation is clearly visible when we compare plots in the first column in Fig.~\ref{Fig6}a. All of them correspond to the interaction strength $g=2.0$ and only for the state $|{\bf 3};{\bf 1}\rangle$ a very slow current of particles through the barrier is observed. 

Let's now discuss the dynamical properties of the system initially prepared in states containing opposite-spin pair, $|{\bf 2};{\bf P}\rangle$ and $|{\bf P};{\bf 2}\rangle$. In these cases, in the non-interacting limit, due to the Pauli exclusion principle, only two particles from opposite wells are allowed to tunnel. Initially, these particles occupy opposite wells and different single-particle levels. From this perspective, the dynamics of the system is exactly equivalent to the dynamics of two-particle initial states $|{\bf 1}^*;{\bf 1}\rangle$ and $|{\bf 1};{\bf 1}^*\rangle$ discussed briefly in Section~\ref{Sec4}. However, along with increasing interactions this correspondence become less visible and the dynamics is rather similar to the cases $|{\bf 1};{\bf P}\rangle$ and $|{\bf P};{\bf 1}\rangle$ with $N=3$ particles discussed previously. It is clear when dependences of the decomposition numbers $\kappa$ are compared (see the right panel in Fig.~\ref{Fig5}c and Fig.~\ref{Fig6}b). In both cases, for a quite large range of interactions dependences are smooth and monotonic and only if interactions are strongly attractive systems enter to regions of unpredictable dynamics caused by contributions of highly-excited many-body states. 

A dynamical distinctness between the separated initial states $|{\bf N_\downarrow};{\bf N_\uparrow}\rangle$ and the states with paired fermions $|{\bf 2};{\bf P}\rangle$ and $|{\bf P};{\bf 2}\rangle$ is even more clear when instead of occupations a dynamical creation of entropies is compared (shown in Fig.~\ref{Fig7}). In the case of states containing localized pairs, for all repulsions and not too strong attractions ($g\gtrsim -1.5$) the entropy is rather produced successively and monotonically in time. Contrary, in the case of separated states, the entropy is rapidly created just after the initial moment and then it slightly oscillates around the average value. In the latter case, for specific values of interactions, inter-component correlations are built much stronger. It is accompanied by a very rapid flow of particles through the inter-well barrier and coefficient $\kappa$ becomes peaked.

Finally, let us note that the dynamical properties of the system are even more interesting when the state with two pairs of opposite-spin fermions $|{\bf P};{\bf P}\rangle$ is considered. This state is particularly interesting since, due to a perfect mirror symmetry of the state, macroscopic current through the barrier cannot be noticed for any interactions, {\it i.e.}, the occupations $N_{L\sigma}(t)$ and $N_{R\sigma}(t)$ remain constant in time and equal to $1$. However, it does not mean that there are no any dynamical consequences of interactions. It is clear that also in this case the inter-component correlations are quickly built after the initial moment (see Fig.~\ref{Fig7}) and entanglement entropy smoothly and monotonically increases along with interactions. Any resonant magnifications are not present.

\section{Dynamics of more than four particles system}
\begin{figure}
\centering
\includegraphics{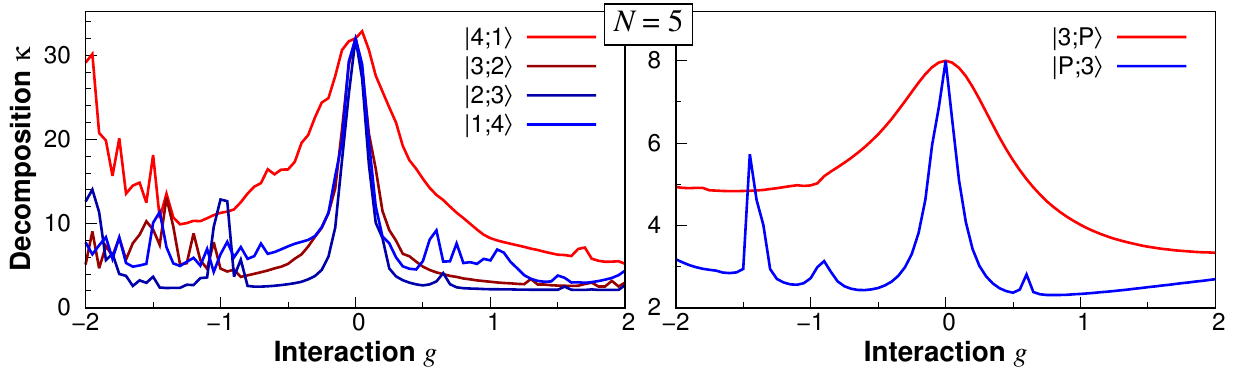}
\includegraphics{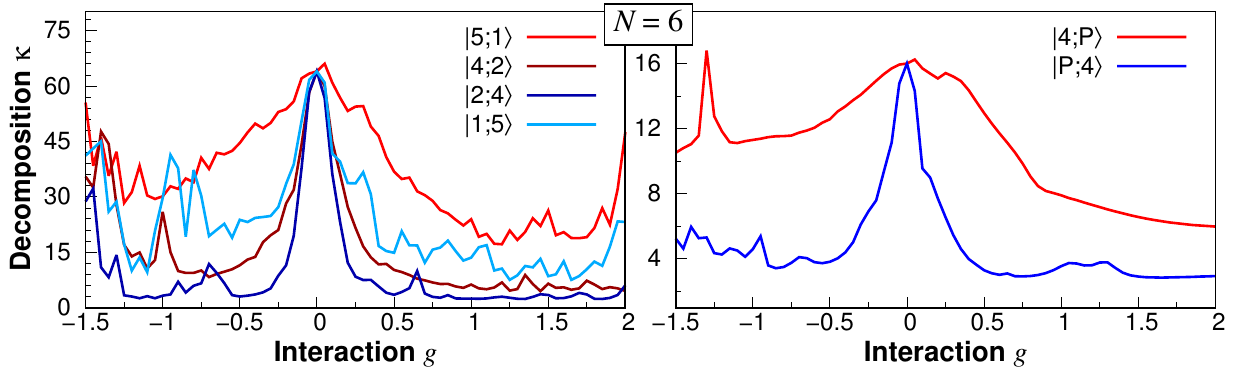}
\caption{Effective number of many-body eigenstates $\kappa$ contributing to different initial states as a function of interactions for systems containing $N_\downarrow+N_\uparrow=5$ (top) and $N_\downarrow+N_\uparrow=6$ (bottom) particles. Interaction strength $g$ is expressed in dimensionless units related to a spin-$\downarrow$ (light) particle, $(\hbar^3\Omega/m_\downarrow)^{1/2}$. \label{Fig8}}
\end{figure}
Analysis of the dynamical properties of the system containing up to four particles shows a consistent picture in the interaction regime $|g| \lesssim 1.5$ within which the tunneling dynamics is only insignificantly influenced by couplings to higher orbitals of the double-well. Beyond this interaction regime, the impact of excited single-particle modes is gradually reflected in the dynamics. This dependency is more pronounced for systems with a larger number of particles and leads to the appearance of regions of unpredictable dynamics. In the region of weak interactions, one can distinguish two qualitatively different behaviors depending on a structure of the initial state (separated clouds or localized pair of opposite-spin fermions). This simplified and phenomenological analysis suggests that similar behavior should be present also for systems containing a larger number of particles (of course in appropriately narrower interaction range). To verify this hypothesis we performed a detailed numerical analysis of systems up to six particles. Indeed, by calculating the decomposition coefficient for different initial states we argue that also these systems have very similar properties to systems with a smaller number of particles. However, due to the increasing richness of the many-body spectra, regions of unpredictable are more noticeable. As an example, in Fig.~\ref{Fig8} we show results for several representatives. 

Further increasing of the number of particles is straightforward, however, it should be done with care. In fact, for a larger number of particles, a whole picture of well-localized single-particle left/right orbitals breaks down. Moreover, from the experimental point of view, it becomes not possible to prepare the system with well-separated components. The critical number of particles, of course, depends on the barrier height and in our case $N_\downarrow$ and $N_\uparrow$ is around five. 

\section{Conclusion}
We have thoroughly investigated the unitary dynamics of mass-imbalanced two-flavors fermions in the double-well potential with a total number of particles up to six. Starting with a minimal system of just two particles we study the dynamics of other many-particle systems to understand the role of the initial state, interactions, mass-imbalance, and the quantum statistics. In particular, we have focused on the dynamical properties of two classes of initial states which can also be easily prepared in the contemporary cold-atom experiments. 

For the two-particle system, the tunneling through the central barrier is generically suppressed when inter-particle interactions are enhanced (repulsive as well as attractive). However, for some particular values of interactions, when the average energy of the initial state is close to eigenenergies of the many-body Hamiltonian, we observe resonant acceleration in tunneling through the barrier. In these resonant points, we observe a significant increase of the number of many-body eigenstates contributing to the state evolution. The resonance points are characterized by a specific avoided crossings in the many-body spectrum of the system which can be viewed as a consequence of coupling between between the center-of-mass and the relative motion degrees of freedom triggered by the anharmonicity of the confining potential.
 
Beyond the two-particle system, in the presence of interactions, the qualitative picture outlined above remains valid. However, due to a much more complicated structure of the many-body spectrum, resonances are not well-isolated and the dynamics become highly unpredictable. By monitoring the evolution of the single-particle von Neumann entropy, we argue that in regions of a strong suppression of the density flow through the barrier, quantum correlations are also built much slower during the evolution. 

Considering a different number of particles, we found that the dynamical properties of the system depend on the structure of the initial state, rather than on the number of particles in individual components. For example, whenever opposite-spin components are initially localized in opposite wells, we distinguish two different regimes of interactions. First, the region of repulsive and not too strong attractions (in our case $g\gtrsim- 1.5$) where suppression of the dynamics monotonically depends on interaction strength. Second, the regime of strong attractions where resonant enhancements of tunneling appear. In contrast, in the case of initial states in which opposite-spin pair is solely localized in selected well, dependence on interaction strength is very predictable and it is accompanied by a very smooth response of the decomposition factor $\kappa$ (except sporadically present resonant crossings).

Finally, let us once more emphasize that in our present analysis we consider a simplified model relying on the assumption that the two fermionic components feel the same external trapping frequencies. Although the model does not treat the problem from the most general perspective, we expect that it explains the most important phenomena appearing -- tunneling resonances, oscillation frequencies, suppressed tunneling regions, {\it etc.}. Using a more general model may change the results only on a quantitatively level (shifting of positions of resonances or regions of suppressed tunnelings). In other words, the present model can provide valuable insight towards some general models that could certainly be studied in the future.
  
\section{Acknowledgments}
This work was supported by the (Polish) National Science Center Grant No. 2016/22/E/ST2/00555. Numerical calculations were partially carried out in the Interdisciplinary Centre for Mathematical and Computational Modelling, University of Warsaw (ICM), under Computational Grant No. G75-6. TS is grateful to the FoKA community for many inspiring discussions.

\section{References}
\bibliographystyle{iopart-num}
\bibliography{mybibDW}

\end{document}